\documentclass[preprint]{aastex}

\usepackage{psfig}

\received{} 
\accepted{} 
\journalid{}{} 
\articleid{}{} 
\shortauthors{Lentz et al.}
\shorttitle{Metallicity in Type I\lowercase{a} Supernovae}

\newcommand{\feff}{$^{54}$Fe}
\newcommand{\snia}{SNe~Ia} 
\newcommand{\phoe}{{\tt PHOENIX}} 
\newcommand{\kmps}{km~s$^{-1}$}

\begin{document}

\title{Metallicity Effects in NLTE Model Atmospheres of Type~I\lowercase{a}
Supernovae}

\author{Eric~J. Lentz, E.~Baron, David Branch}

\affil{Department of Physics and Astronomy, University of Oklahoma, 440 West Brooks,
Norman, OK 73019-0225 \email{baron,branch,lentz@mail.nhn.ou.edu}}

\author{Peter~H. Hauschildt}

\affil{Department of Physics and Astronomy \& Center for Simulational Physics,
University of Georgia, Athens, GA 30602}

\email{yeti@hal.physast.uga.edu}
\author{and}

\author{Peter~E. Nugent}

\affil{Lawrence Berkeley National Laboratory, Berkeley, CA 94720}

\email{penugent@lbl.gov}

\begin{abstract}

We have calculated a grid of photospheric phase atmospheres of Type~Ia
supernovae (\snia) with metallicities from ten times to one thirtieth the
solar metallicity in the C+O layer of the deflagration model, W7. We
have modeled the spectra using the multi-purpose NLTE model-atmosphere
and spectrum-synthesis code, \phoe. We show models for the epochs 7, 10,
15, 20, and 35 days after explosion. When compared to observed
spectra obtained at the approximately
corresponding epochs these synthetic spectra fit reasonably well.   The
spectra show variation in the overall level of the UV 
continuum with lower fluxes for models with higher metallicity in the 
unburned C+O layer. This is consistent with the classical surface cooling
and line blocking effect  due to metals in the outer
layers of C+O.
The UV features also move consistently to the blue with higher
metallicity, demonstrating that they are forming at shallower and faster
layers in the atmosphere.  The potentially most useful effect
is the blueward movement of the Si~II feature at 6150~\AA\ with
increasing   C+O layer metallicity.  We also demonstrate the more complex
effects of  metallicity variations by modifying the \feff\ content of the incomplete burning zone
in W7 at  maximum light.   We briefly address some shortcomings of the W7
model when compared to observations.
Finally, we identify that the split in the Ca H+K feature produced in
W7 and observed in some SNe Ia is due to a blending effect of
Ca~II and Si~II and does not necessarily represent a complex abundance or
ionization effect in Ca~II.
\end{abstract}


\section{Introduction}

Type Ia supernovae (\snia)  are recognized for their near uniformity as
standard candles. This has led to the use of \snia\ in cosmology
\citep{brancharaa98,garnetal98,riess_scoop98,perletal99}.  The development of
empirical 
calibrations between peak brightness and light curve shape
\citep{philm15,hametal96a,rpk96,perlasi97,perlq097,tb99},
have enhanced the
usefulness of \snia\ as distance  
indicators. Nevertheless theoretical modelers have yet to agree on the
source of these apparently systematic variations.  A primary concern
is the evolutionary lifetime of \snia\ progenitors and the possibility
of significant deviations of distant \snia\ from their well observed
counterparts in the local galactic neighborhood. 
If galactic chemical evolution occurs slowly, then more distant SNe
arise from a younger, metal poor population. On the other hand,
the metallicity variations in the local sample may already span the
range of the entire observational sample.

We probe the possible effects of progenitor metallicity
variations on the observed spectra, by modifying the
parameterized  
deflagration model, W7 \citep{nomw7,nomw72}.  Using base fits to
observations~(\S~\ref{basemod}) we have scaled all elements heavier than
oxygen in the unburned C+O layer of W7 to simulate the effects of
various metallicities in the progenitor system. \citet{hwt98}
explored this question by modifying the
pre-explosion metal content of a particular \snia\ model and noted the
differences in final composition, light curves, and spectra.
They found that the composition of the partially burned
layers 
of the ejecta yielded larger quantities of \feff. A similar effect can be 
seen in the lowered \feff\ abundance when W7 is calculated using a pure 
C+O mixture without other metals \citep{iwamoto99}. 
We have modified the
\feff\ abundance of the partially burned layers of W7, to replicate this
effect. While \citet{hwt98} focused mainly on the effects of
metallicity variations on the light curve and energetics, here we
concentrate exclusively on its effects on the observed spectra,
particularly at
early times where the formation of the spectrum occurs in the unburned C+O
layer, which  is most sensitive to initial progenitor metallicity.

Our computational methods are given in \S~\ref{methods}.  In
\S~\ref{zco} we show the effects of metallicity modification of the C+O
layer on the synthetic spectra at each epoch. In \S~\ref{feint} we show the effects of
\feff\ on day 20 and 35 model spectra. In \S~\ref{siii} we discuss the
evolution of the Si~II feature at 6150~\AA.

\section{Methods\label{methods}}

\subsection{Abundance Modification}

To model the effects of metallicity on \snia, we have
modified the base W7 model at several epochs. W7, while clearly not the
complete model for \snia, is a good starting point for
spectral modeling. The composition
structure of W7 as a function of velocity does reasonably well
in reproducing the observed spectra of \snia\ 
\citep{harkness91a,harkness91b,branch81b85,nughydro97}. 
By making separate calculations to consider the direct (C+O metallicity)
and the indirect explosive nucleosynthetic consequences (\feff\ enhancement),
our computational
methods allow us to probe these effects of progenitor metallicity
independently. \citet{hwt98} changed the composition before
computing the hydrodynamics and explosive nucleosynthesis of the model.
This gives them consistent nucleosynthesis and energetics within the context
of their chosen input physics.  However, it is difficult to separate effects
in spectra that arise from different consequences of the
initial metallicity variation.

 To modify the metallicity
of the unburned 
C+O layer we have scaled the number abundances of elements
heavier than oxygen 
in the velocity range 14800--30000~\kmps\ by a constant factor $\zeta$,
such that for all species $i$ heavier than oxygen, the new number abundance
$n_i'$, is given by 
\[
 n_i' = \zeta n_i. \]
The mass fractions are then renormalized,
\[
  \sum X_i = 1,\]
in each layer.  For all epochs
we have used C+O metallicity factors, $\zeta$, of 10, 3, 1, 1/3, 1/10,
and 1/30. 
These effects are discussed in detail in \S~\ref{zco}.

The nucleosynthesis results of \citet{hwt98} show that
modifying the metallicity of the
progenitor white dwarf changes the quantity of \feff\ produced in the
incomplete burning zone in a manner approximately proportional to the
metallicity. 
 This is due to excess neutrons in the progenitor over a pure C+O
mixture, particularly $^{22}$Ne \citep[see for example][and references
therein]{arnettbk96,nomw7,iwamoto99}.  To simulate this effect in W7, we have
scaled the \feff\ abundance
in the incomplete burning zone of W7, 8800--14800~\kmps,
in the same manner as the C+O layer metallicity,
\[ n_{^{54}{\rm Fe}}' = \xi n_{^{54}{\rm Fe}}. \]
The mass fractions are then renormalized
in each layer.  We have used
\feff\ abundance factors, $\xi$, of 3, 1, 1/3, 1/10, and 1/30 for the epochs 20 and 35
days after explosion.  For the remaining epochs we have only computed
models for 1/10 \feff\ abundance ($\xi=0.1$).  These effects are discussed in detail in
\S~\ref{feint}.

Finally, we simulate the full effects of progenitor metallicity on the
output spectra, by combining the two effects, C+O layer metallicity
and incomplete burning zone \feff\ abundance, into a single series
of models.  For each model we have used the same factor for both C+O
metallicity and \feff\ abundance ($\xi=\zeta$).  These factors are the same as for the
\feff\ abundance modifications above.  These effects are discussed in detail in
\S~\ref{feint}.

\subsection{Computational Methods of \phoe\label{phx}}

The calculations were performed using the multi-purpose stellar
atmospheres program \phoe~{\tt 9.1} 
\citep{hbjcam99,bhpar298,hbapara97,phhnovetal97,phhnovfe296}.
\phoe\ solves the radiative
transfer equation along characteristic rays in spherical symmetry
including 
all special relativistic  effects.  The non-LTE (NLTE) rate
equations for several ionization states are solved including the
effects of ionization from non-thermal electrons from the $\gamma$-decay
energy of the $^{56}$Ni core.  The atoms and ions calculated in NLTE
are: He~I~(11 levels), He~II~(10), C~I~(228), O~I~(36), Na~I~(3),
Mg~II~(18), Si~II~(94), S~II~(85), Ti~II~(204), Fe~II~(617), and
Co~II~(255). Each model atom includes primary NLTE transitions, which
are used to calculate the level populations and opacity, and weaker secondary NLTE
transitions which are 
are included in the opacity and implicitly affect the rate equations
via their effect on the solution to the transport equation.  In
addition to the NLTE transitions, a 
number of LTE line opacities for atomic species not treated in NLTE 
are treated with the equivalent two-level atom source
function, using a thermalization parameter, $\alpha =0.1$, as in
\citet{nughydro97} for LTE lines for all models in this paper.  The
atmospheres are iterated to energy balance in the co-moving frame,
\begin{eqnarray}
\frac{\gamma}{r^2}\frac{\partial (r^2 H)}{\partial r} &+&\nonumber\\
 \gamma\beta
\frac{\partial J}{\partial r}& +& \gamma\left[\frac{\beta}{r}(3J-K)\right. +\nonumber\\
&&\left.\gamma^2 \frac{\partial\beta}{\partial r}(J+K+2\beta
 H)\right]\nonumber\\
&=& \frac{\dot{S}}{4\pi},\nonumber\\
\end{eqnarray}
where, $\dot{S}$ is the deposited energy from radioactive decay, 
$\gamma$ and $\beta$ have their usual meaning and $c=1$.  This
equation neglects the explicit effects of time dependence in the
radiation transport equation, however the term on the right hand side
implicitly includes these effects.
The
models are parameterized only by the day,  which determine the radii
and amount of
radioactive decay, and by the luminosity parameter, $\eta$, which is defined
as,
\[
L_{bol}=\eta L_{\gamma}^{abs},
\]
where $L_{\gamma}^{abs}$ is the instantaneous deposition of
radioactive decay $\gamma$-ray energy.  The deposition of
$\gamma$-rays is determined 
by solving a grey transport equation
\citep{nugphd,nughydro97,sw84} as a function of time. This approach
has been shown to be of sufficient accuracy for our purposes
when compared to detailed Monte-Carlo calculations
\citep{swsuthhark95}.
We use two different
inner boundary 
conditions for our models.  For days 7, 10, and 15, which are optically 
thick at the core, we use a
diffusive inner boundary condition.  For days 20 and 35 which have
lower total optical depth we  use a `nebular' inner
boundary.  This boundary condition takes the downward intensity of any
ray and assumes that it passes unimpeded through the core of the
atmosphere, becoming the upward intensity on the other side, with only
the necessary Lorentz transformation. Additional details
on the use of \phoe\ in modeling \snia\ atmospheres can be found in
\citet{nugphd}.

\subsection{Baseline Models\label{basemod}}

To perform numerical experiments on the metallicity effects in the
spectra of \snia, we need reasonable base models for each epoch.  For each
day, an observed spectrum of SN~1994D was fitted with a W7 model of
appropriate age.  Several models were computed with various luminosity
parameters, $\eta$, to fine tune the spectral shape. A separate paper
fitting W7 to SN~1994D photospheric era spectra (Lentz et al. in
preparation) shows the comparisons to observations. 
\citet{hofkhoklc96} find $\eta$ (their  $\tilde Q$) in the range $0.7
< \eta < 1.8$. When $\eta > 1$ stored radiative energy is being
released and when $\eta < 1$ radiative energy is being stored.

The maximum light
(epoch day 0 for the observations) fit for SN~1994D to W7 (with $\eta =
1.0$) at day 20 after the explosion is the same as in
\citet{nughydro97}.  For the two model fits at days 10 (which we compare
to day -9 with
respect to B maximum) and 15
(compared to day -4 with
respect to B maximum) for 
W7 we have used luminosity parameters of $\eta=0.4$
and $\eta=0.8$ 
respectively. For these two models the fits are generally good, but with
all of the pre-maximum spectra, the red edge of the Si~II
feature at 6150~\AA\ does not extend far enough to the red.
For model day 7 (fit to day -12) we chose the
the model with luminosity $\eta=0.1$. This model fits the luminosity and
most features well; however, like the day 10 models 
the Ca~H+K feature not  extend blueward enough when compared to the
earliest observations.
\citet{hatano94D99} have modeled the same spectra
and found that the Ca~H+K feature requires ionized calcium with
velocities up to $\sim 40000$~\kmps.  We have conducted experiments
extending the density profile of W7.  This did not provide enough
optical depth in the Ca~H+K feature to affect the line profile.  Neither
do our models reproduce the secondary feature at 4700~\AA\ that
\citet{hatano94D99} 
ascribe to Fe~II absorption in the C+O layer.  For our
post-maximum model, day 35 (15 days after maximum), we fit the observation
with a model with luminosity, $\eta=1.5$.  This model fits generally;
however, the large feature at 4800~\AA\ (probably Fe~II) is not seen in
the observations.  The observed Na~I D-line feature is missing,
probably due to deficiencies in the sodium model atom, which has already
been improved in the latest version of \texttt{PHOENIX}. Here, we are
interested in small differential effects so it was essential that all
models be run with the code version frozen.

\section{Metallicity of Unburned C+O Layer\label{zco}}

%By conducting numerical experiments we can
%separate different input effects and examine their effects on the emergent 
%spectra independently.  
The simplest effect of metallicity on the spectral formation in \snia\ is the
change in the metal content of the unburned C+O layer neglecting
changes in the density structure and deeper layers.  In this section
we examine these effects 
independently of other effects of progenitor metallicity on \snia.\
Since ongoing supernova searches are expected to discover SNe early,
and early spectra probe the outermost layers only, this approach is
sensible and yields physical insight that is somewhat model independent.

\begin{figure} 
\begin{center}
\leavevmode
%\epsscale{0.8} 
%\plotone{d07zall.eps}
\psfig{file=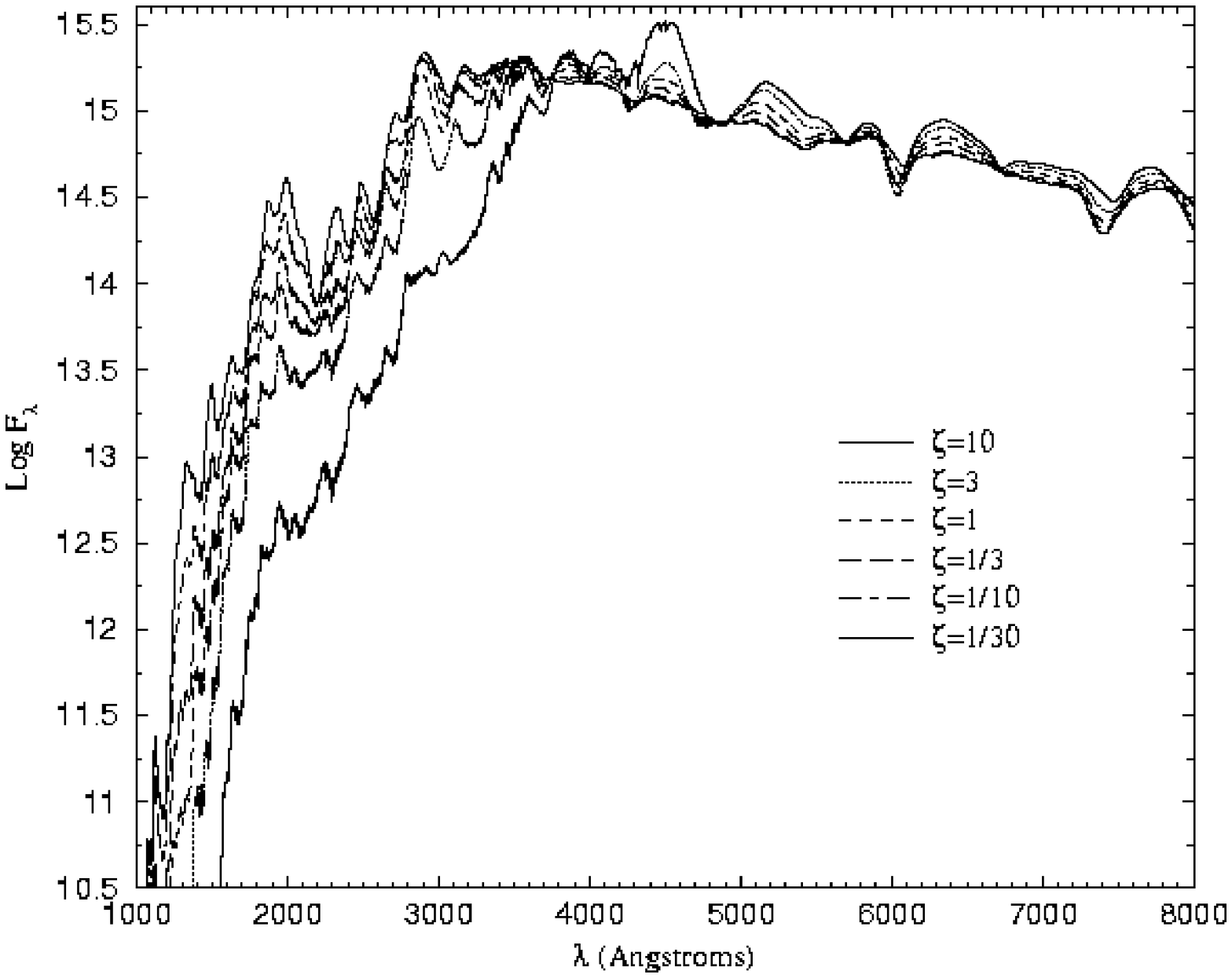,width=14cm,angle=270}
\caption{\label{fig:07zall} Models with various metallicities in C+O
layer at 7 days after explosion.  Thick solid denotes 10 times the
normal C+O metallicity, thick dotted---3 times, short dashed---normal,
long dashed---1/3, dot-dashed---1/10, and thin solid---1/30.}
\end{center}

\end{figure}

\begin{figure} 
\begin{center}
\leavevmode
%\epsscale{0.8} 
%\plotone{d07zuv.eps}
\psfig{file=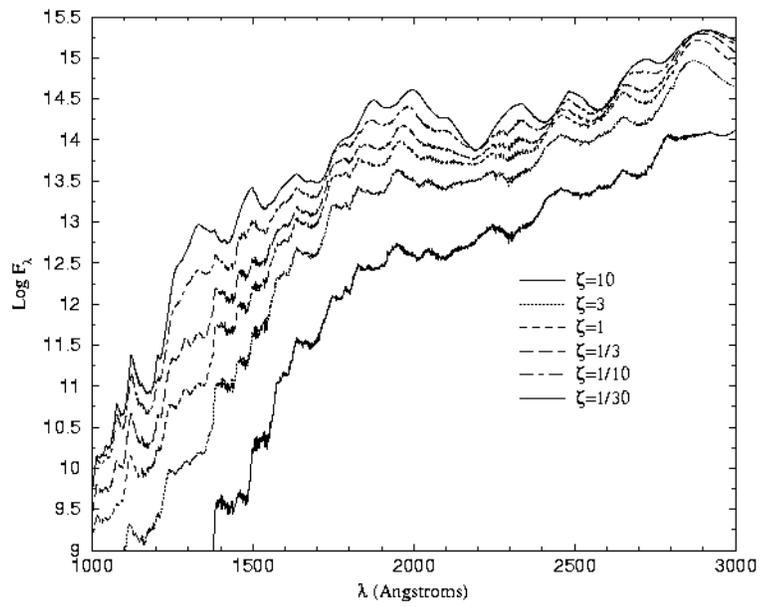,width=14cm,angle=270}
\caption{\label{fig:07zuv} Expansion of ultraviolet region of
Figure~\ref{fig:07zall}}
\end{center}
\end{figure}

\begin{figure} 
\begin{center}
\leavevmode
%\epsscale{0.8} 
%\plotone{si.eps}
\psfig{file=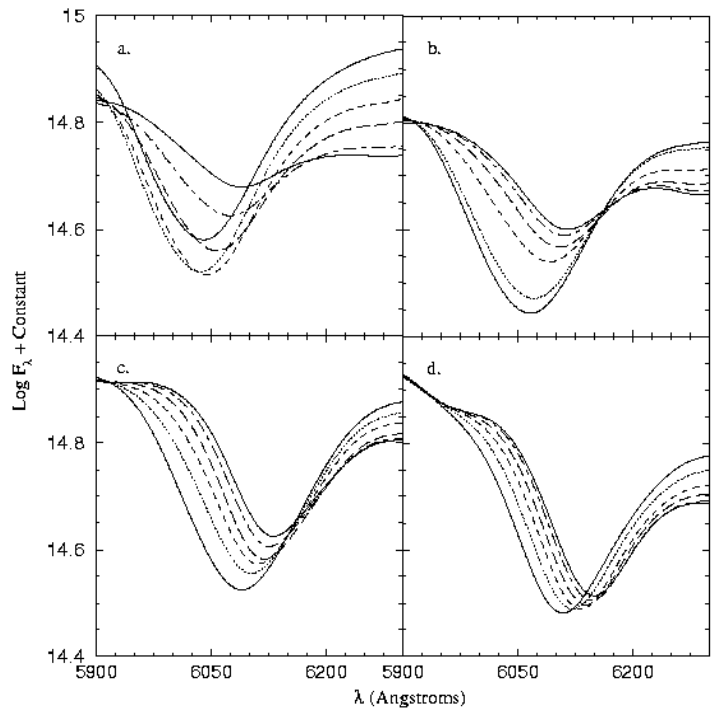,width=14cm,angle=270}
\caption{\label{fig:sigrid} Expansion of Si~II 6150~\AA\ region of
a. Figure~\ref{fig:07zall} (Day 7), b. Figure~\ref{fig:10zall} (Day 10), 
c. Figure~\ref{fig:15zall} (Day 15), and d. Figure~\ref{fig:20zall} (Day 20).}
\end{center}
\end{figure}

\begin{figure} 
\begin{center}
\leavevmode
%\epsscale{0.8} 
%\plotone{d10zall.eps}
\psfig{file=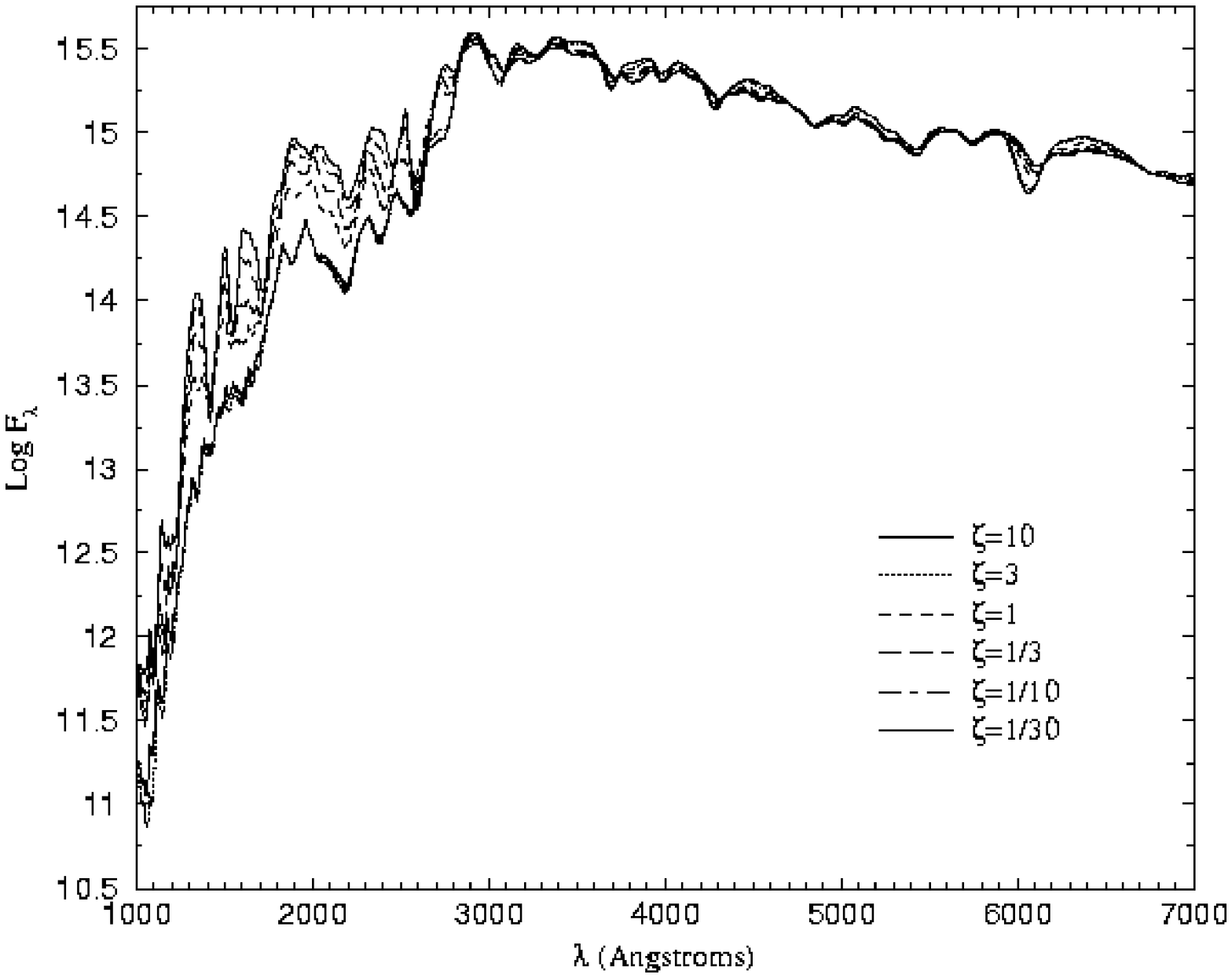,width=14cm,angle=270}
\caption{\label{fig:10zall} Models with various metallicities in C+O
layer at 10 
days after explosion.  Labels are the same as
Figure~\ref{fig:07zall}.} 
\end{center}
\end{figure}

\begin{figure} 
\begin{center}
\leavevmode
%\epsscale{0.8} 
%\plotone{d15zall.eps}
\psfig{file=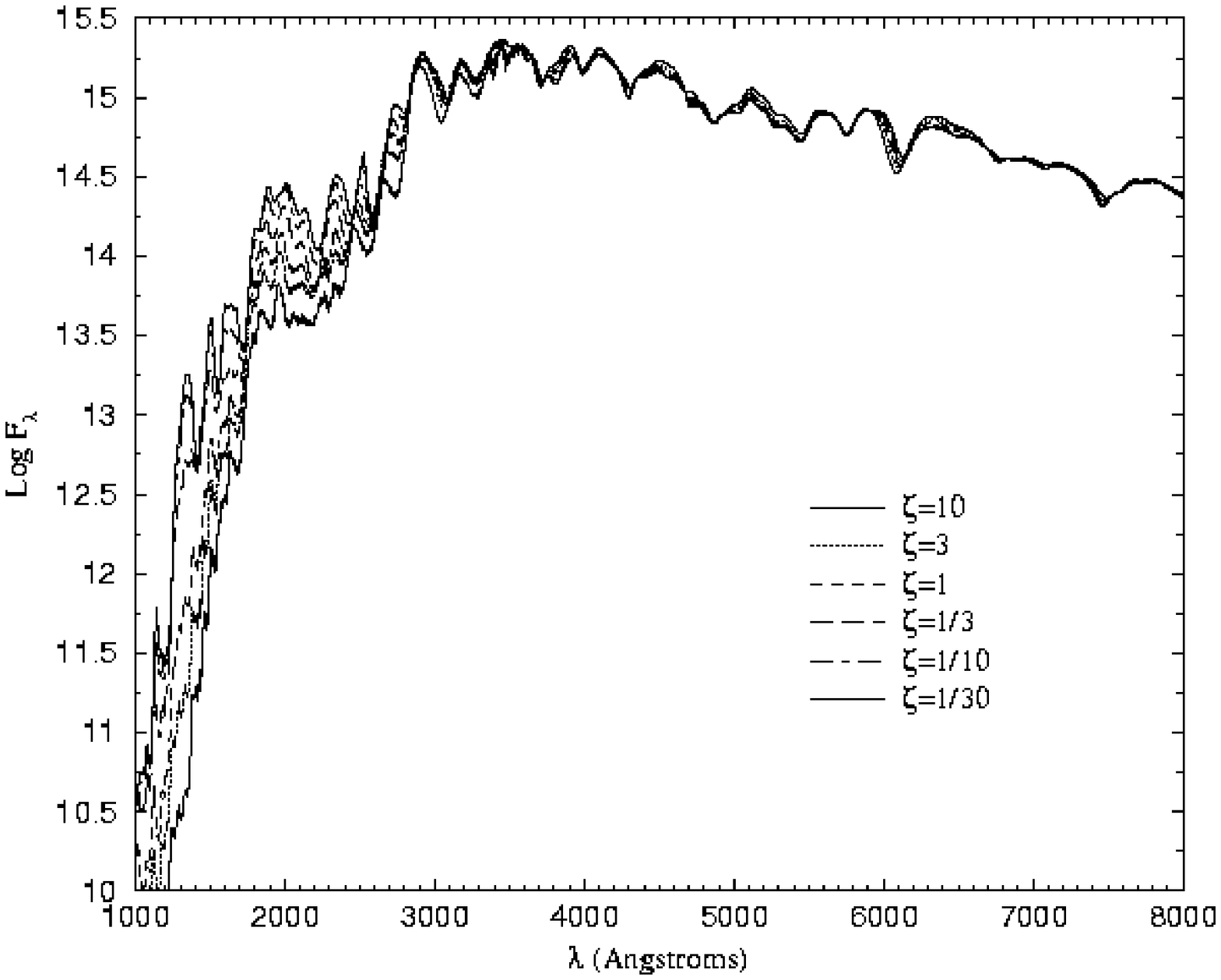,width=14cm,angle=270}
\caption{\label{fig:15zall} Models with various metallicities in C+O
layer at 15 
days after explosion.  Labels are the same as Figure~\ref{fig:07zall}.}
\end{center}
\end{figure}

\begin{figure} 
\begin{center}
\leavevmode
%\epsscale{0.8} 
%\plotone{d20zall.eps}
\psfig{file=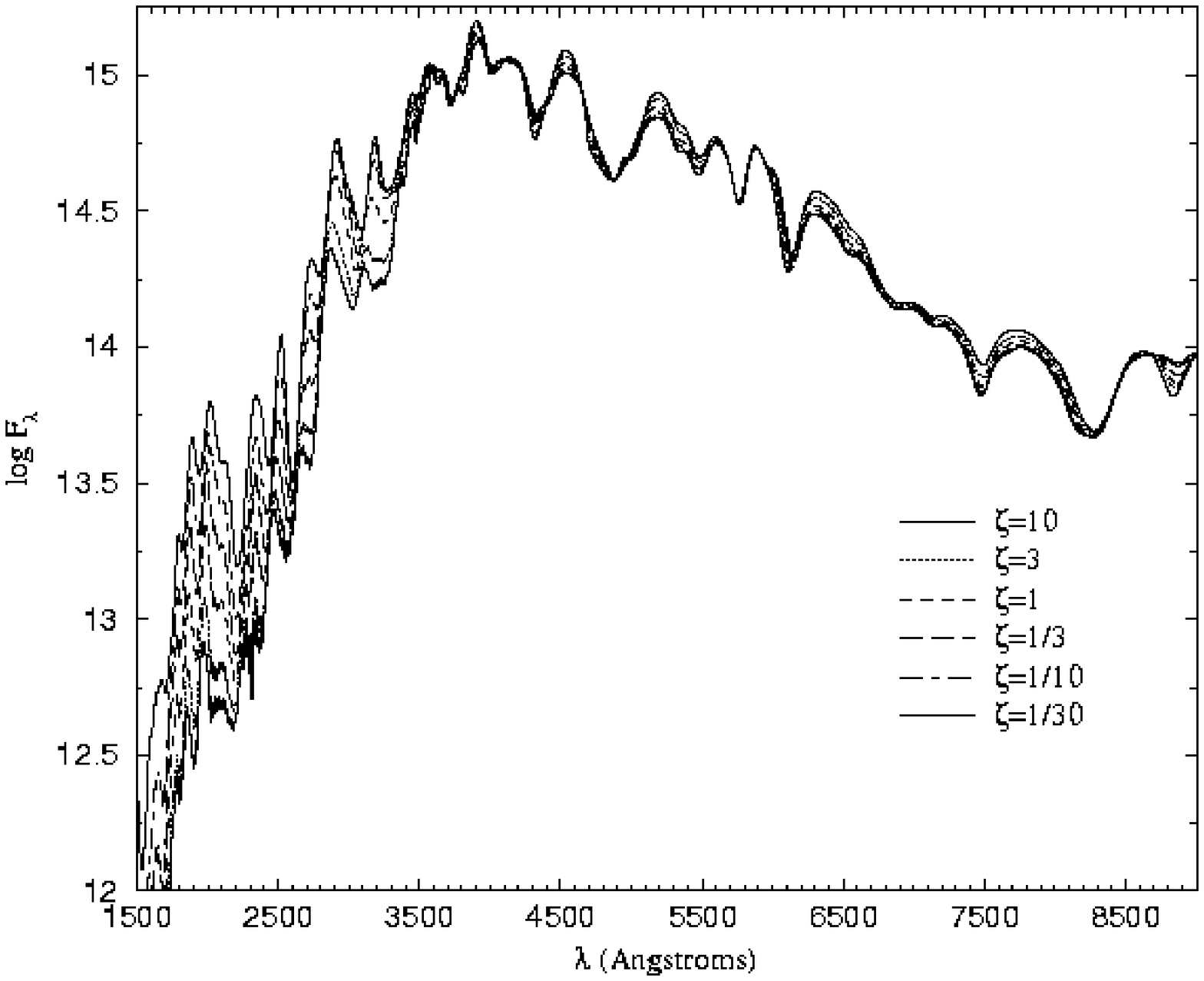,width=14cm,angle=270}
\caption{\label{fig:20zall} Models with various metallicities in C+O
layer at 20 days after explosion.  Labels are the same as
Figure~\ref{fig:07zall}.}
\end{center}
\end{figure}

\begin{figure} 
\begin{center}
\leavevmode
%\epsscale{0.8} 
%\plotone{d35zall.eps}
\psfig{file=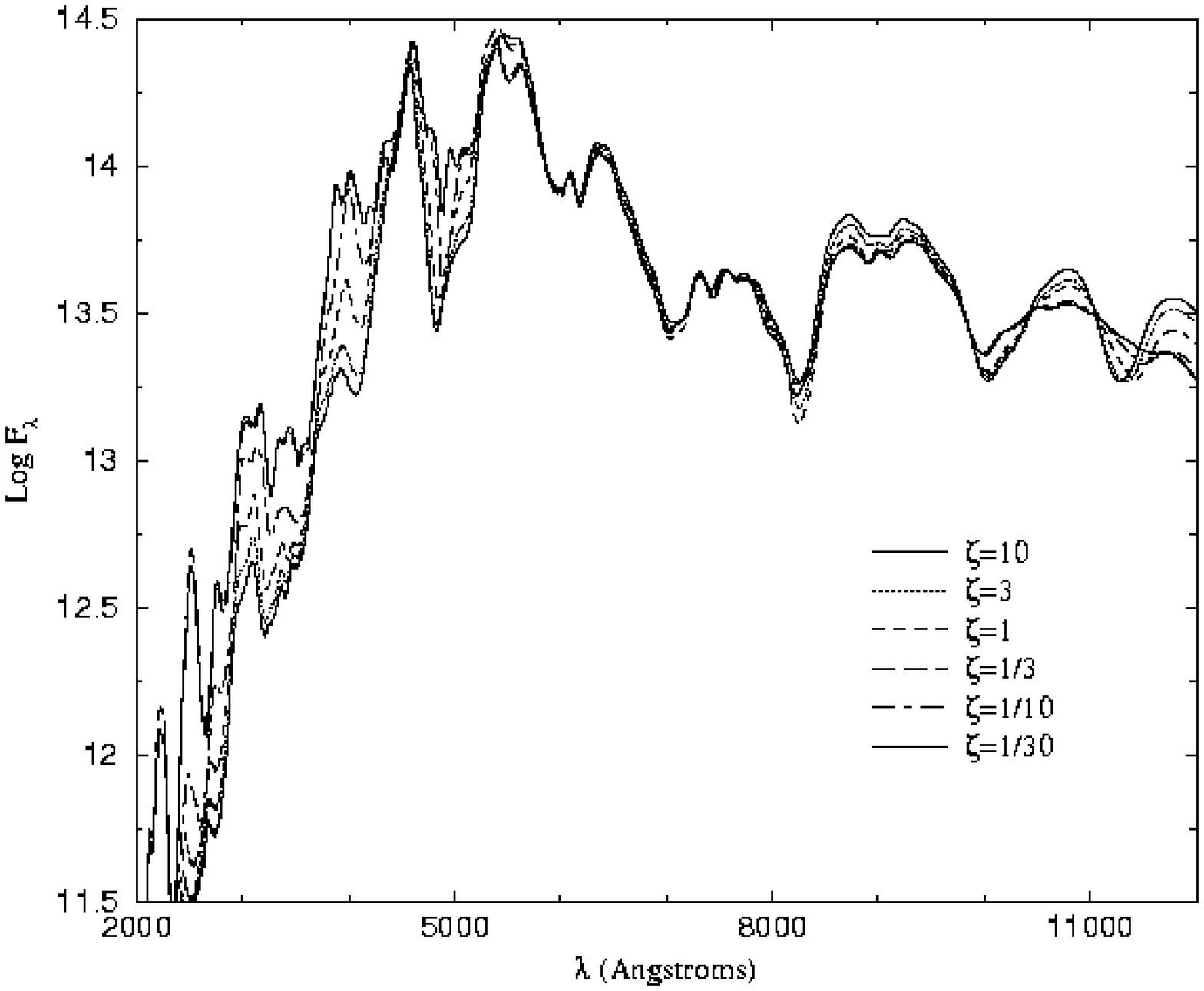,width=14cm,angle=270}
\caption{\label{fig:35zall} Models with various metallicities in C+O
layer at 35 days after explosion.  Labels are the same as
Figure~\ref{fig:07zall}.}
\end{center}
\end{figure}

When we look at the overall UVOIR synthetic spectra
(Figures~\ref{fig:07zall}, \ref{fig:10zall}, \ref{fig:15zall},
\ref{fig:20zall}, and~\ref{fig:35zall}) of the models with variations in
the C+O layer metallicity, we see two consistent and significant effects: shifts in
the UV pseudo-continuum level (expanded view for day 7 in Figure~\ref{fig:07zuv})
%\ref{fig:10zuv}, \ref{fig:15zuv}, \ref{fig:20zuv}, and~\ref{fig:35zuv})
and variations in the Si~II line at 6150~\AA\ (expanded views in
Figure~\ref{fig:sigrid}).

The general effect in the UV is the increase in the UV
pseudo-continuum level with decreasing metallicity.  Simultaneous is
the redward (blueward) shift of most UV features with decreasing
(increasing) metallicity.  In the UV, the line-forming region is in
the C+O layer.  As the metallicity decreases, the line forming region
must reach deeper into the atmosphere to have the same line opacity,
resulting in smaller line velocities.  Modification of the C+O layer
metal abundance gives a classic surface cooling effect, lower
temperatures for higher metallicity.  The higher temperatures of the
lower metallicity C+O atmospheres give higher thermal fluxes, moving
the UV pseudo-continuum higher with lower metallicity.  The surface
cooling and the resulting shifts in UV pseudo-continuum are evident at
every epoch. There is the complementary effect of additional metals
increasing the line blocking. We make no attempt to separate these two
effects in this paper.

The Si~II line at 6150~\AA\ (Figure~\ref{fig:sigrid}) shifts blueward
with increases in 
metallicity for epochs through day 20.  These shifts demonstrate that
some line formation in this feature takes place in the C+O layer.  The
earlier epochs show large variations in the total depth of the feature
which implies that the line forms less in the incomplete burning zone
with its large, unchanging silicon abundance, and more in the C+O
layer where the silicon
abundance changes. At later epochs these conditions are reversed,
resulting in smaller changes with C+O layer metallicity
variation. These effects are discussed further in \S~\ref{siii}.

The Mg~II~``h+k'' feature at 2600~\AA\ (Figure~\ref{fig:07zuv}) does not move
to the blue or red as the metallicity varies.  
The decrease in Mg~II h+k feature strength with increasing metallicity
is caused by the increasing UV line blanketing from background line 
opacity in the C+O layer.  The Mg~II absorption occurs mostly in the deeper,
partially burned layer that is highly enriched in magnesium.  We have confirmed
this hypothesis by calculating diagnostic output spectra without using
any background opacity \citep[see][for a discussion of the method]{b94i2}.
These diagnostic spectra suggest that this feature  (and the feature
near 2800~\AA) may actually be due to a complicated blend
of Mg~II UV 1-4 transitions: h+k (UV 1) $\lambda 2798$, (UV 2) $\lambda
2796$, (UV 3) $\lambda 2933$, and (UV 4) $\lambda 2660$.

\subsection{Day 7\label{z07}}

\begin{figure} 
\begin{center}
\leavevmode
%\epsscale{0.8} 
%\plotone{d07zopt.eps}
\psfig{file=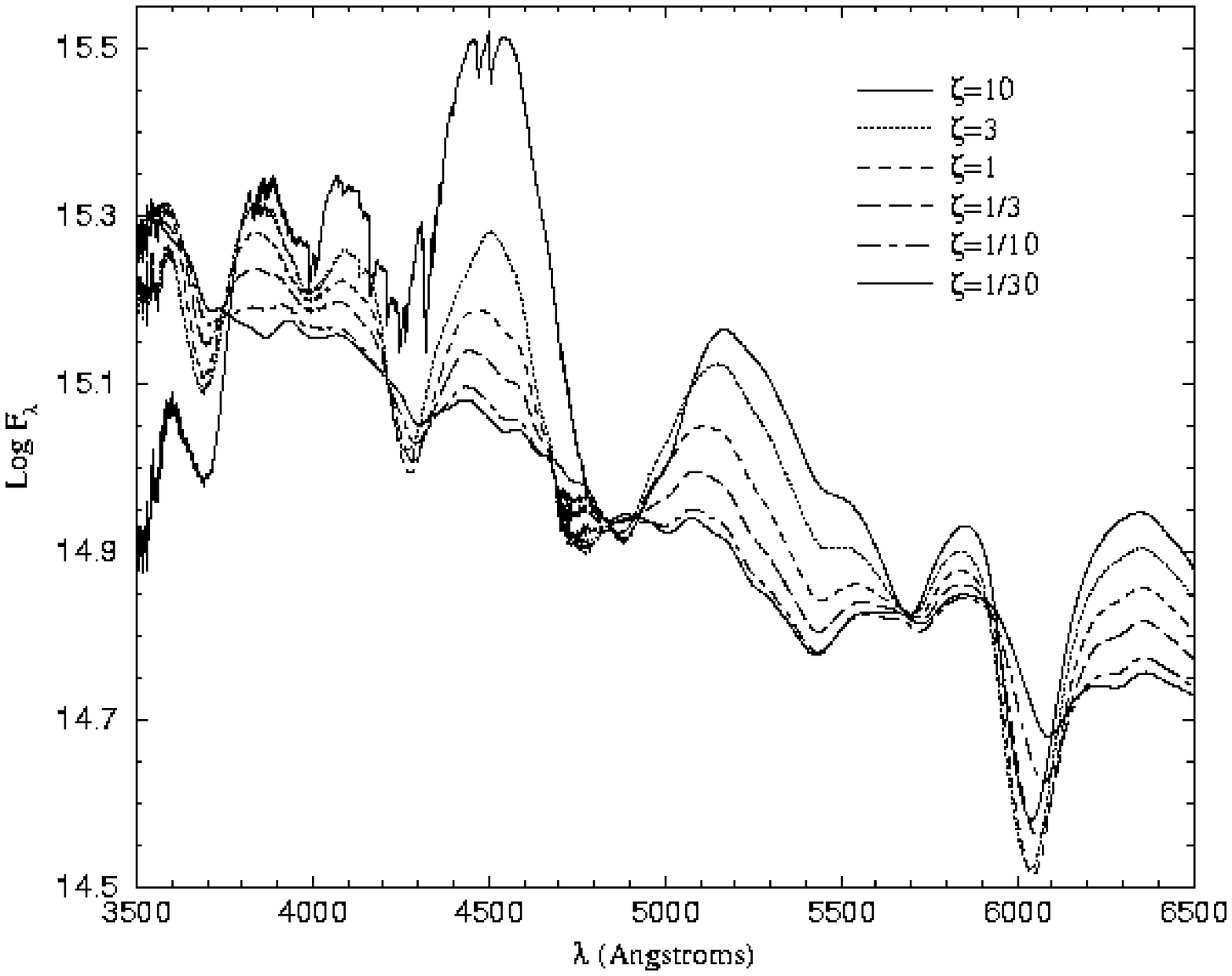,width=14cm,angle=270}
\caption{\label{fig:07zopt} Expansion of optical region of
Figure~\ref{fig:07zall}}
\end{center}
\end{figure}

\begin{figure} 
\begin{center}
\leavevmode
%\epsscale{0.8} 
%\plotone{d07zir.eps}
\psfig{file=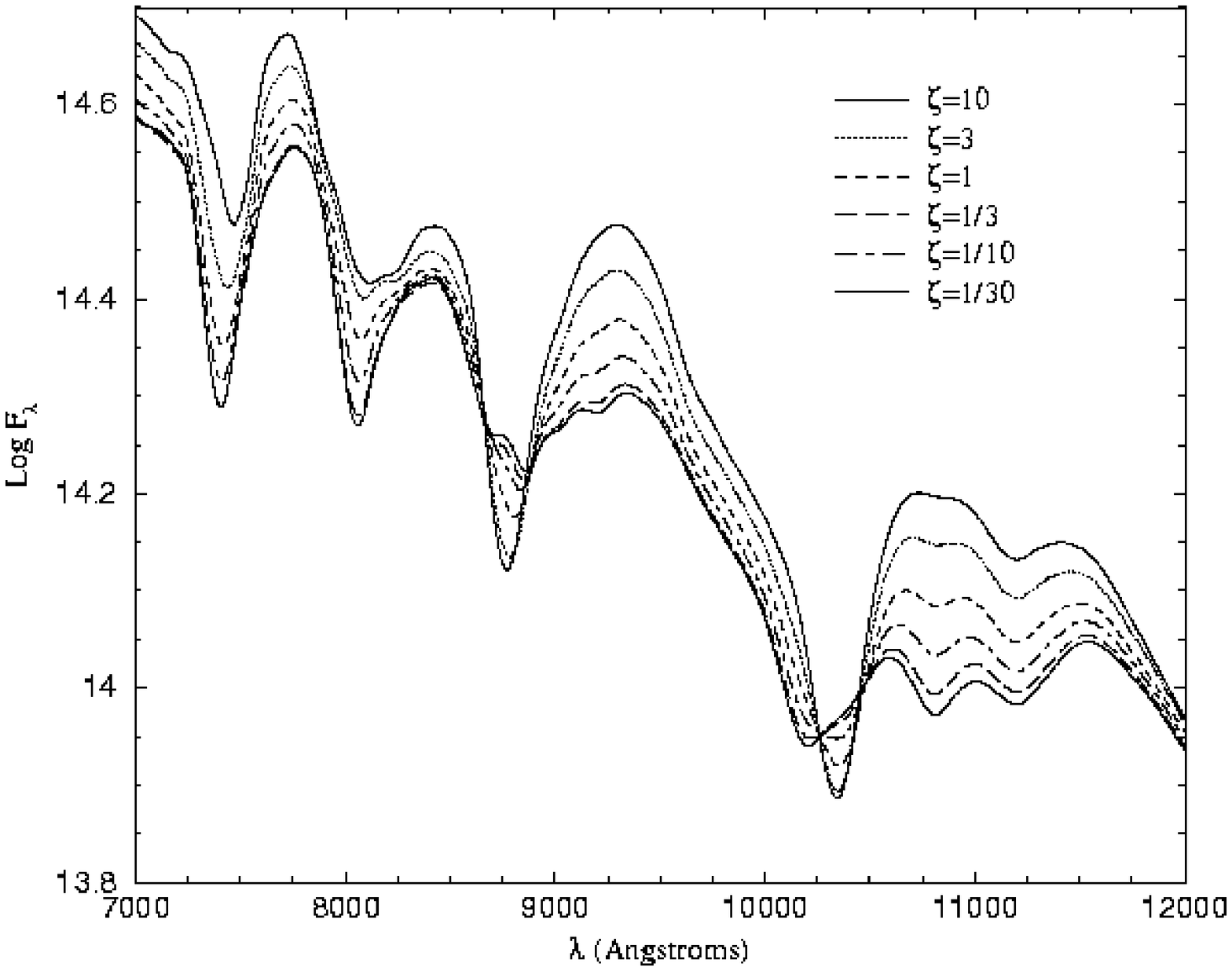,width=14cm,angle=270}
\caption{\label{fig:07zir} Expansion of infrared region of
Figure~\ref{fig:07zall}}
\end{center}
\end{figure}

Our grid of synthetic spectra of W7 on day 7 with C+O metallicity
variations are shown in Figure~\ref{fig:07zall}. The case $\zeta=10$
is likely extreme and is shown for illustrative purposes only, we
don't consider it further at this epoch. The UV spectra
(Figure~\ref{fig:07zuv}),
show variations in the  pseudo-continuum  due to surface cooling, line
blocking, and
feature shifts 
due to changes in the  feature formation depth (velocity).
The pseudo-continuum level in the optical and IR 
vary with metallicity.  Lower metallicities have lower pseudo-continuum
fluxes.  Figures~\ref{fig:07zopt} and~\ref{fig:07zir} display
optical and infrared spectra  which illustrate this effect more
clearly. This is due to backwarming.
  Figure~\ref{fig:07zopt} shows that the
line features at 3650~\AA\ (Ca~II H+K), 4250~\AA\ (Fe~II, Mg~II,
\&~Si~III), 4700~\AA\ (Fe~II), and 5650~\AA\ (Si~II) nearly disappear
as the metallicity drops to 1/30 of normal.  The 4700~\AA\ Fe~II
feature is weak and doesn't extend far enough to the blue in the base
model fits (\S~\ref{basemod}).  Increasing the metallicity by even a
factor of 10 does not extend the feature blueward enough (to sufficiently
large velocities) to match the observations.  The same is true of the
Ca~II H+K lines. This indicates that more mass extending to
higher velocities is needed to fit this feature.  The Si~II feature at
6150~\AA\ (Figure~\ref{fig:sigrid}a) shows the blueward movement of
the feature minimum and blue-edge wall with increasing metallicity.
The slope of the red edge of the feature changes and at the lowest
metallicities the feature is not strong enough to form the red
emission wing of the P Cygni feature.  The infrared spectra
(Figure~\ref{fig:07zir}) show several features that weaken with lower
metallicity, Mg~II $\lambda9226$~\AA\ and
Si~I complexes at $\lambda10482$
and $\lambda10869$~\AA\ \citep[cf.][]{millard94i99}, as well as
two features that {\it strengthen} with lower metallicity.  These two
O~I features (7400~\AA\ and 8150~\AA) weaken as the oxygen abundances
decreases in favor of
higher Z metals.

\subsection{Day 10\label{z10}}

\begin{figure} 
\begin{center}
\leavevmode
%\epsscale{0.8} 
%\plotone{d10zopt.eps}
\psfig{file=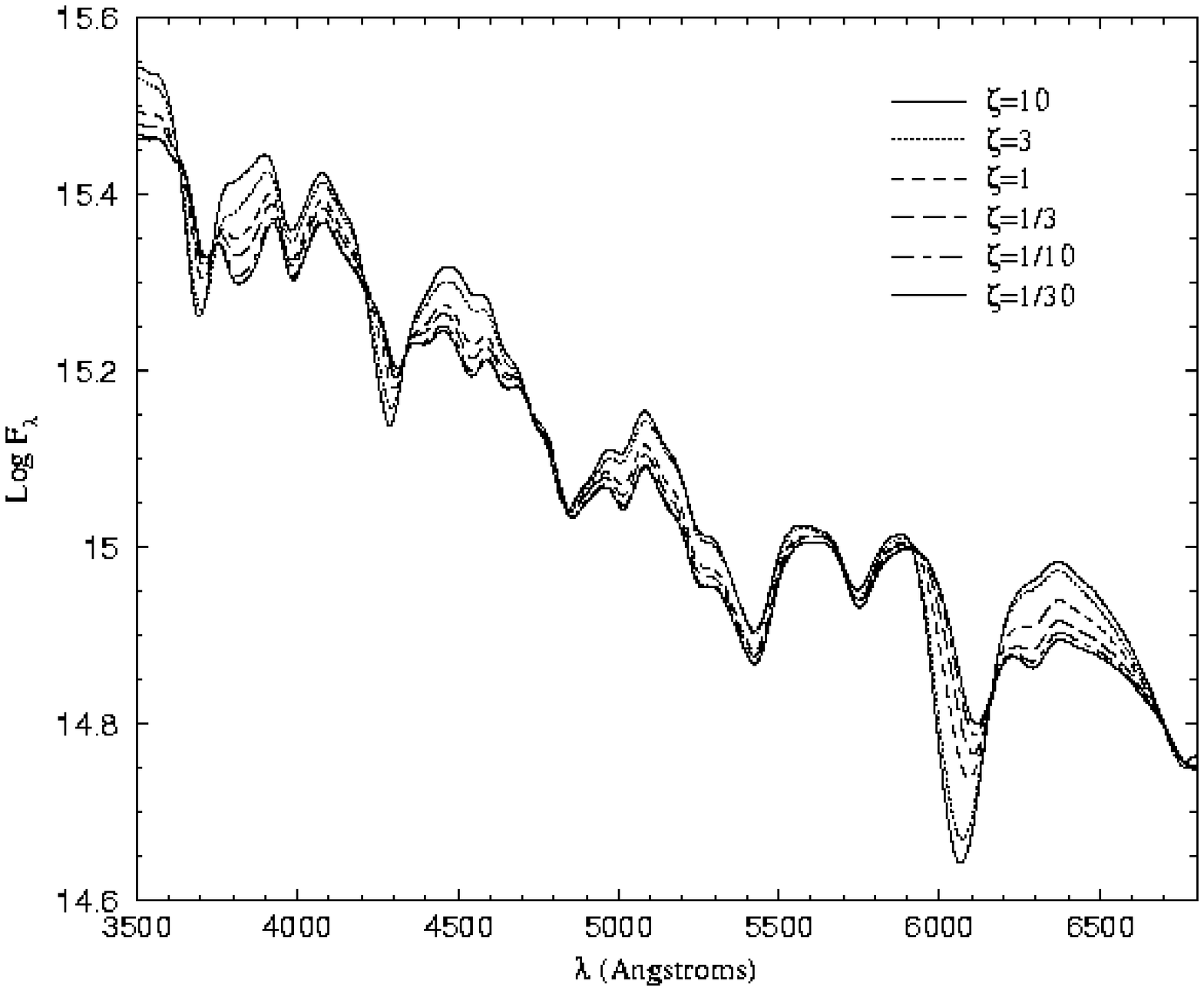,width=14cm,angle=270}
\caption{\label{fig:10zopt} Expansion of optical region of
Figure~\ref{fig:10zall}}
\end{center}
\end{figure}

Our grid of synthetic spectra of W7 on day 10 with metallicity
variations in the C+O layer (Figure~\ref{fig:10zall}) display
similar variations 
to those in the day 7 spectra (Figure~\ref{fig:07zall}). The UV variations
 again show the blue-shifted features and lower continua with
increasing metallicity.  The optical pseudo-continuum effects
(Figure~\ref{fig:10zopt}) are still present, but smaller. We again see
lines that deepen with increased metallicity such as 3650~\AA\ (Ca~II
H+K), 4250~\AA\ (Fe~II, Mg~II, \&~Si~III), and 6100~\AA\ (Si~II), but
these effects are less dramatic than at day 7.  The Si~II
feature (Figure~\ref{fig:sigrid}b) still shows the changing depth and
slopes of the line edges with metallicity, but, at day 10 the depth of
the Si~II feature in the lowest metallicity model is larger than in
the same model on day 7, relative to the highest C+O layer metallicity
in the respective epoch.  These effects on the optical lines indicate that line
formation for certain strong lines is now taking place in layers below
the C+O layer.  The effects of line formation below the C+O layer
becomes more important as the supernova atmosphere expands and becomes
less opaque.

\subsection{Day 15\label{z15}}

\begin{figure} 
\begin{center}
\leavevmode 
%\epsscale{0.8} 
%\plotone{d15zblue.eps}
\psfig{file=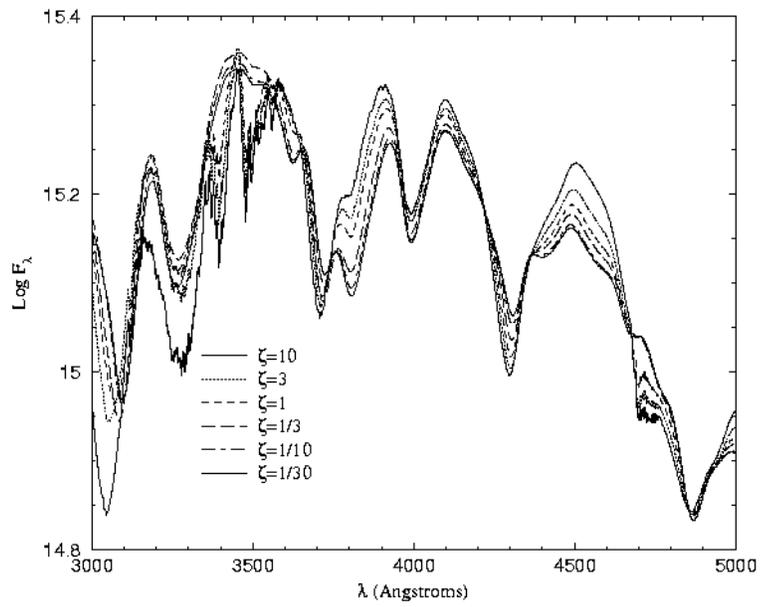,width=14cm,angle=270}
\caption{\label{fig:15zblue} Expansion of blue regions of
Figure~\ref{fig:15zall}}
\end{center}
\end{figure}

\begin{figure} 
\begin{center}
\leavevmode
%\epsscale{0.8} 
%\plotone{d15znir.eps}
\psfig{file=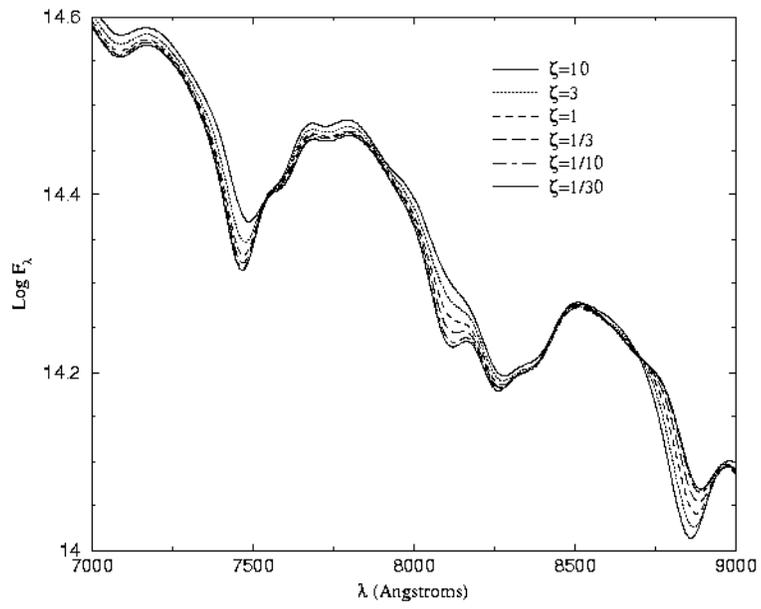,width=14cm,angle=270}
\caption{\label{fig:15znir} Expansion of near infrared region of
Figure~\ref{fig:15zall}}
\end{center}
\end{figure}

Our grid of synthetic spectra of W7 on day 15 with C+O metallicity
variations (Figure~\ref{fig:15zall}) are again  similar to those in the
day 7 and 10 spectra (Figures~\ref{fig:07zall}
\&~\ref{fig:10zall}).  The decreasing effects of C+O layer metallicity
are apparent.  The UV pseudo-continuum variation with metallicity
remains strong. The optical (Figure~\ref{fig:15zblue}) and 
the near infrared (Figure~\ref{fig:15znir}) show the backwarming
optical/IR pseudo-continuum flux effect which occurred in the optical
at day 10, and extended 
well into the IR at day 7. Features which increased in strength with
increasing C+O metallicity are a blend of Fe~II, Mg~II, \&~Si~III at 4250~\AA,
the multi-species blend at 3300~\AA, Mg~II at 8850~\AA, and an
unidentified feature at 10400~\AA\ (possibly Si~I).  
 The feature at 3300~\AA\ is a blend of 
weak lines that forms in the C+O layer. 
Figure~\ref{fig:15zblue} illustrates the blue shifting of
features as 
changes in metal content move the depth (and thus
velocity)  
of line formation.

The two component feature at 3700--3900~\AA\ is usually labeled Ca~II
H+K. \citet{nughydro97} have identified this `Ca split' as arising
from a blend of Ca~II H+K with Si~II $\lambda 3858$.  We have
calculated a series of diagnostic spectra at this epoch using the same
temperature structure, but without background opacity.  We have
confirmed the identification of the blue wing with the Si~II line.
When spectra are calculated without any other line opacities, Ca~II
forms a pair of features. %, each wider than the separate parts in the observed feature.
The blue Ca~II absorption forms from the $\lambda 3727$ line, the red absorption
from the H+K lines. The Si~II feature falls on the
peak between 
the two Ca~II absorptions.  As the Si~II feature strengthens, the flux
displaced by its absorption `fills-in' the Ca~II H+K absorption
creating a `split'. With enough Si~II only the blue absorption feature
remains, while the red feature becomes an inflection.  The lack of a
split in some \snia\ indicates that either the Si~II feature is weaker
or the Ca~II H+K is stronger, preventing the formation of a split.
This helps to illustrate that in supernovae of all types, features are
often the result of more than one multiplet, or even ionic species.

The Si~II feature at 6150~\AA\ (Figure~\ref{fig:sigrid}c) shows much
smaller, but still significant, effects due to C+O metallicity.  The
blue edges are parallel and the red edges nearly so.  The depth
contrast is now much smaller than before.  This is strong evidence
that the feature is forming primarily below the C+O layer at this epoch.
Figure~\ref{fig:15znir} 
shows the displacement of oxygen by metals in the O~I lines at 7450~\AA,
8100~\AA, and 8250~\AA, the latter two of which are superimposed on
the stronger Ca~II `IR-triplet' which, like the H+K absorption, forms
in the calcium rich incomplete burning zone and is unaffected by the C+O
metallicity.

\subsection{Day 20\label{z20}}

\begin{figure} 
\begin{center}
\leavevmode
%\epsscale{0.8} 
%\plotone{d20zopt.eps}
\psfig{file=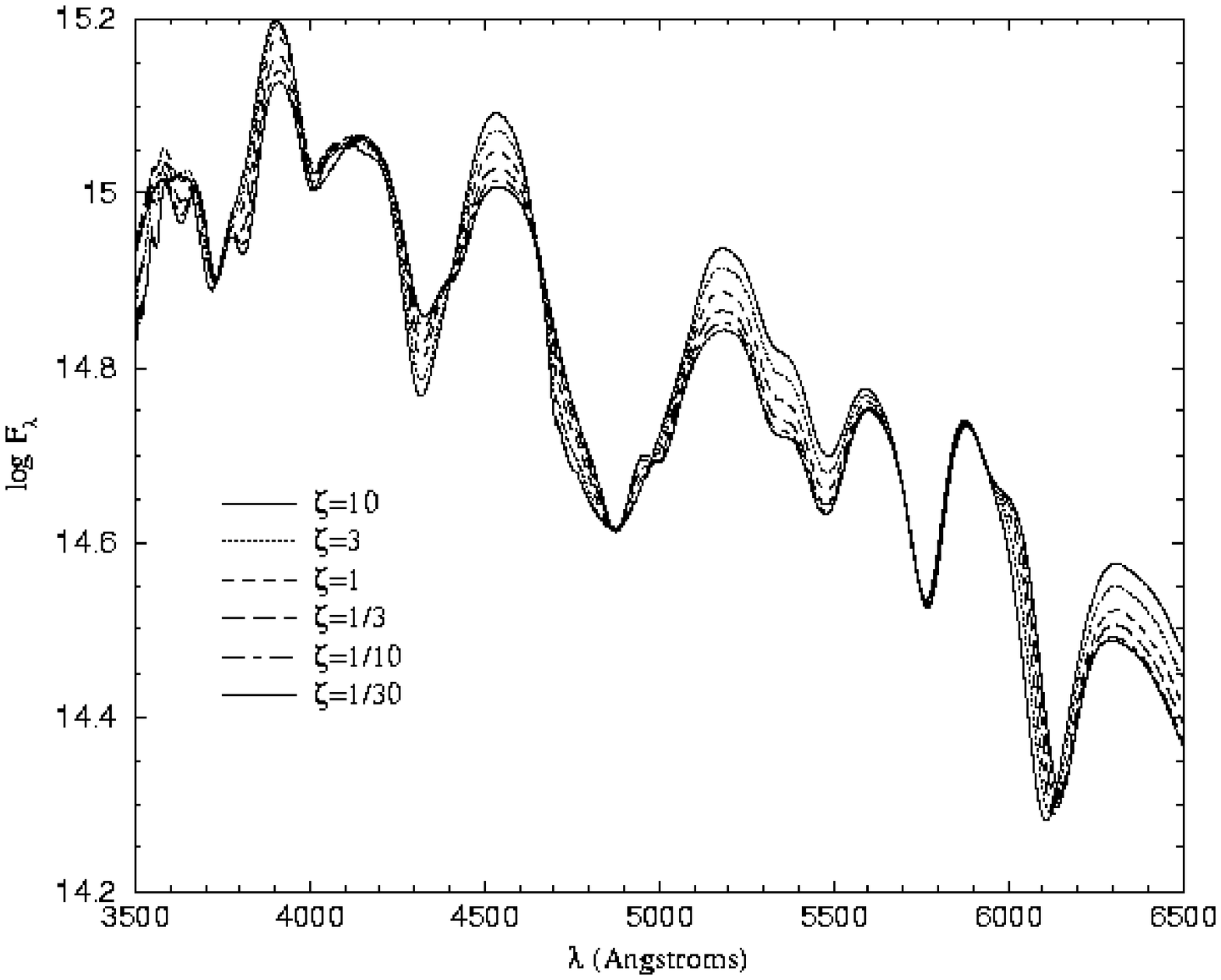,width=14cm,angle=270}
\caption{\label{fig:20zopt} Expansion of optical region of
Figure~\ref{fig:20zall}}
\end{center}
\end{figure}

Our grid of synthetic spectra of W7 on day 20, approximately maximum B
magnitude, with C+O metallicity variations (Figure~\ref{fig:20zall})
shows the continuing reduction in importance
of the C+O layer metallicity as the atmosphere becomes more optically
thin. The UV pseudo-continuum and feature shift
effects are still strong.  The only remaining features which display C+O layer
metallicity effects in the optical (Figure~\ref{fig:20zopt}) are the
Ca~II H+K, the 4250~\AA\ (Fe~II, Mg~II, \&~Si~III), and 6150~\AA\
Si~II features.  As at day 15, the `Ca split' remains, the blue
component increasing in strength and the red component decreasing with
increasing C+O metallicity.  The Si~II feature at 6150~\AA\
(Figure~\ref{fig:sigrid}d) now shows parallel edges (both blue and
red) for the various C+O layer metallicities.  The relatively small
changes in the line strength indicate that the feature forms mostly
in the deeper, silicon rich layers, but some measurable effects due to C+O
layer metallicity are still evident.

\subsection{Day 35\label{z35}}

Our grid of synthetic spectra of W7 on day 35 with C+O metallicity
variations are plotted in Figure~\ref{fig:35zall}.  The UV
features
still show the same pseudo-continuum and line
shifting behavior seen in the previous epochs.  This illustrates that even
at this epoch the pseudo-continuum formation in the UV is still in the unburned
C+O layer. 
The Si~II feature at
6150~\AA\ has only small variations which can not be separated from the
pseudo-continuum effects. It should be noted that the Si~II feature at this
epoch fits the observations rather poorly.

\begin{figure} 
\begin{center}
\leavevmode
%\epsscale{0.8} 
%\plotone{d35zanalysis.eps}
\psfig{file=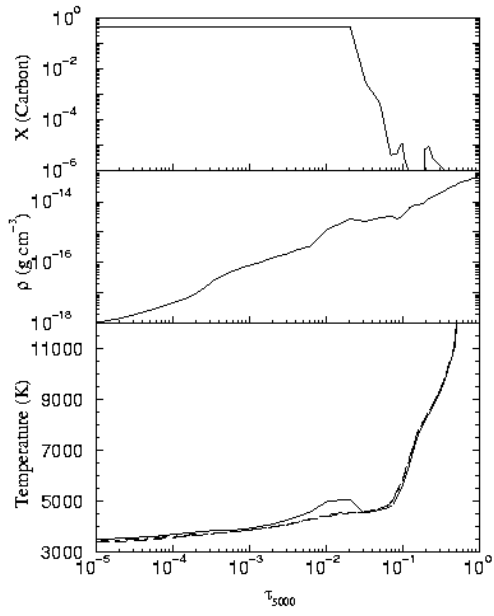,width=14cm,angle=270}
\caption{\label{fig:35zanalysis} Panels from top to bottom display
carbon abundance, density, and temperature profiles for low
metallicity models 
at day 35. The line styles for the temperature
profiles are the same as for the corresponding spectra in Figure~\ref{fig:35zall}.}  
\end{center}
\end{figure}

In the optical  we see the
blue-shifting of the 4800~\AA\ (Fe~II) feature with increasing
metallicity without significant changes to the overall width, shape, or
depth of the feature. This indicates that it forms mostly at deeper
layers, with smaller effects from the C+O layer. 
When looking for a base fit for this epoch, we find that while the line
shapes near 5000~\AA\ are better with somewhat hotter models, those
models had poor overall spectral shape or color. To diagnose this sudden
change in model behavior, we have plotted the temperature profile of the
three models with the lowest metallicity, the density profile, and the
carbon abundance (Figure~\ref{fig:35zanalysis}).  We can see that the
$\zeta=1/30$ model has a definite temperature inversion.  This
inversion corresponds to the position of the density spike in W7
arising from the deflagration wave.  The higher density coincides with 
the change from the more efficient cooling available by intermediate
mass elements to that of the less efficient cooling of C+O
creating the temperature inversion.  In models with
higher metal content the effectiveness of cooling 
by  metals in the C+O layer provides the needed cooling to prevent the
formation of an inversion.

\section{$^{54}$F\lowercase{e} Content of Incomplete Burning Zone\label{feint}}

\subsection{Pre-Maximum Light Epochs}

We have computed models with \feff\ abundance reductions of 1/10 in
the intermediate burning zone and models with the combined 1/10 C+O
metallicity and 1/10 \feff\ abundance reductions for the epochs 7, 10,
and~15 days.  When we compared the \feff\ reduced abundance models with the
related models containing the same C+O metallicities (normal and 1/10
normal, respectively) we found no changes in feature strength.  Some
slight differences in pseudo-continuum levels were seen in the day 10
and~15 comparisons. These may be due to weaker versions of the \feff\
abundance caused backwarming described in \S~\ref{fe20}.  Since each comparison
pair includes two models with different composition in the incomplete
burning zone this may slightly affect the temperature structure.

\subsection{Day 20\label{fe20}}

\begin{figure}
\begin{center}
\leavevmode
%\epsscale{0.8} 
%\plotone{d20feall.eps}
\psfig{file=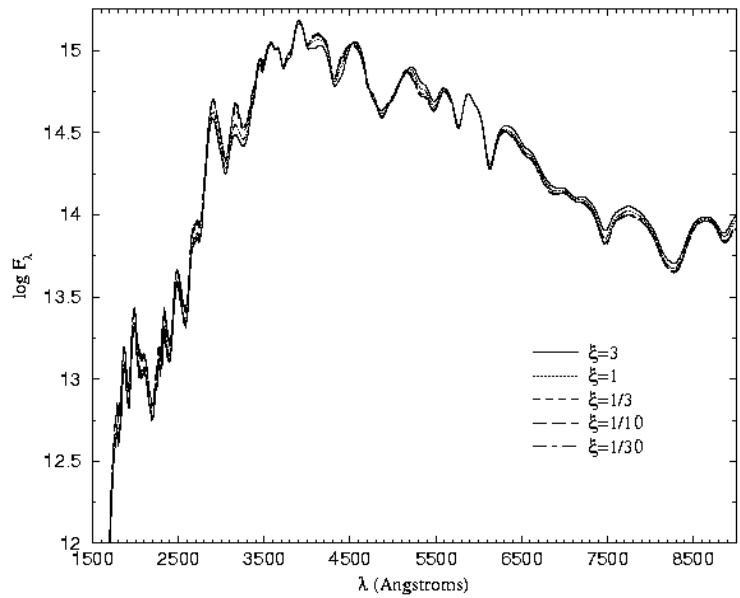,width=14cm,angle=270}
\caption{\label{fig:20feall} Models with various \feff\ abundances in
the incomplete burning layer at 20 days after explosion.  Solid
denotes 3 times the normal \feff\ abundance in the incomplete burning
zone, thick dotted---normal, short dashed------1/3, long
dashed---1/10, and dot-dashed---1/30.}
\end{center}
\end{figure}

Our grid of synthetic spectra of W7 on day 20, approximately maximum B
magnitude, with incomplete burning zone \feff\ abundance variations is shown in
Figure~\ref{fig:20feall}. Some small vertical displacements of the UV
flux,
without blue- or red-shift, can be understood
by the effects of surface cooling/backwarming in the incomplete burning
zone. As intermediate mass elements are replaced by \feff, the larger
line opacity of iron cools the \feff\ rich models.  Since this does not
affect the temperature gradient in the C+O layer, the temperature shift
remains constant throughout the C+O layer. 

\begin{figure} 
\begin{center}
\leavevmode
%\epsscale{0.8} 
%\plotone{d20feopt.eps}
\psfig{file=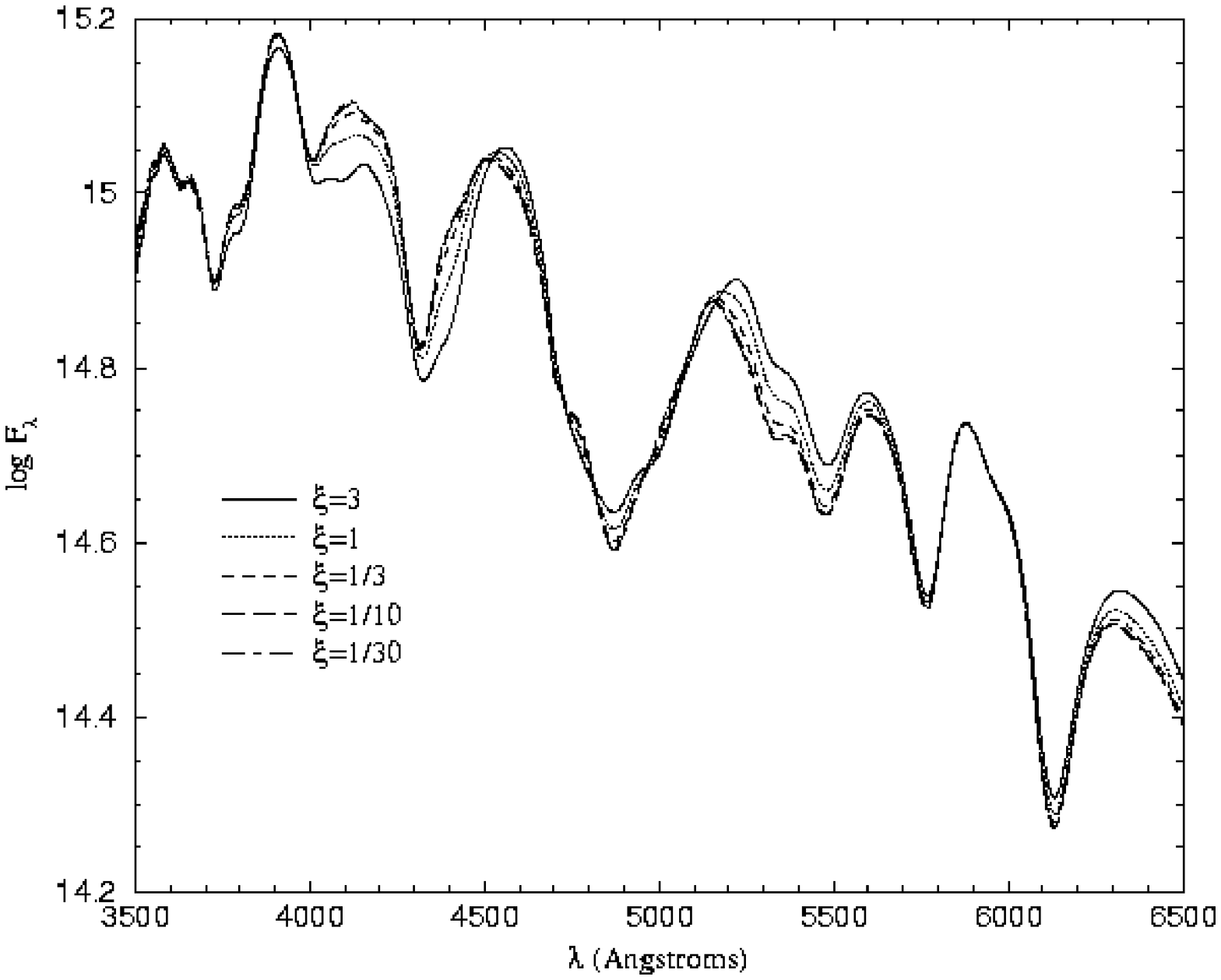,width=14cm,angle=270}
\caption{\label{fig:20feopt} Expansion of optical region of
Figure~\ref{fig:20feall}}
\end{center}
\end{figure}

 In the optical (Figure~\ref{fig:20feopt}) we can
see a few features that {\it decrease} with increasing \feff, such as the
Si~II and~S~II in the 4850~\AA\ feature, the S~II~``W'' at 5300~\AA, and
the Si~II feature at 6150~\AA.  This is caused by
displacement of the species forming the line
by additional \feff.  Several features can be seen to strengthen
with greater \feff\ abundance.  This indicates that they are formed at
least in part 
by iron in the incomplete burning zone. These features included the
feature at 4100~\AA\ that erodes the peak of a neighboring feature and
the red wing of the 4300~\AA\ feature.  

\begin{figure}
\begin{center}
\leavevmode
%\epsscale{0.8} 
%\plotone{d20fezall.eps}
\psfig{file=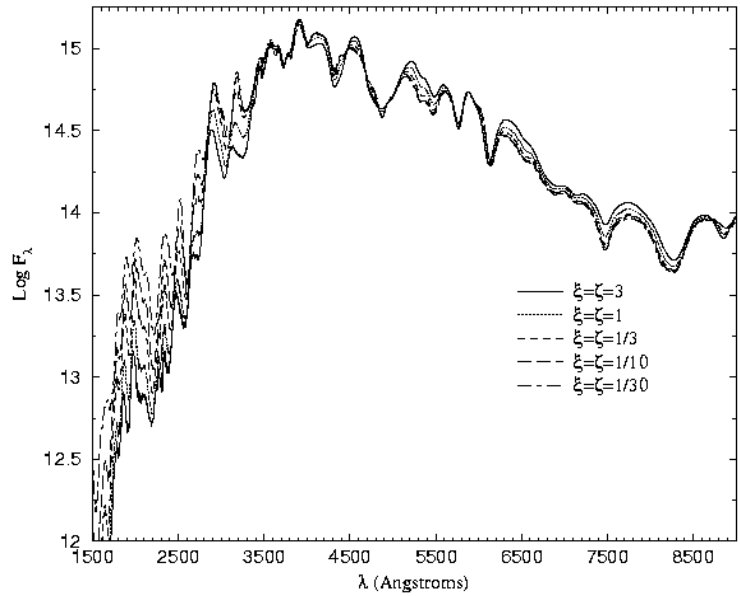,width=14cm,angle=270}
\caption{\label{fig:20fezall} Models with various \feff\ abundances in
the incomplete burning layer and metallicities in the C+O layer at 20
days after explosion.  These models combine the effects in the models
in Figures~\ref{fig:20zall} \&~\ref{fig:20feall}.  Solid denotes 3
times the normal \feff\ abundance in the incomplete burning zone and
C+O layer metallicity, thick dotted---normal, short dashed------1/3,
long dashed---1/10, and dot-dashed---1/30.}
\end{center}
\end{figure}

\begin{figure} 
\begin{center}
\leavevmode
%\epsscale{0.8} 
%\plotone{d20fezopt.eps}
\psfig{file=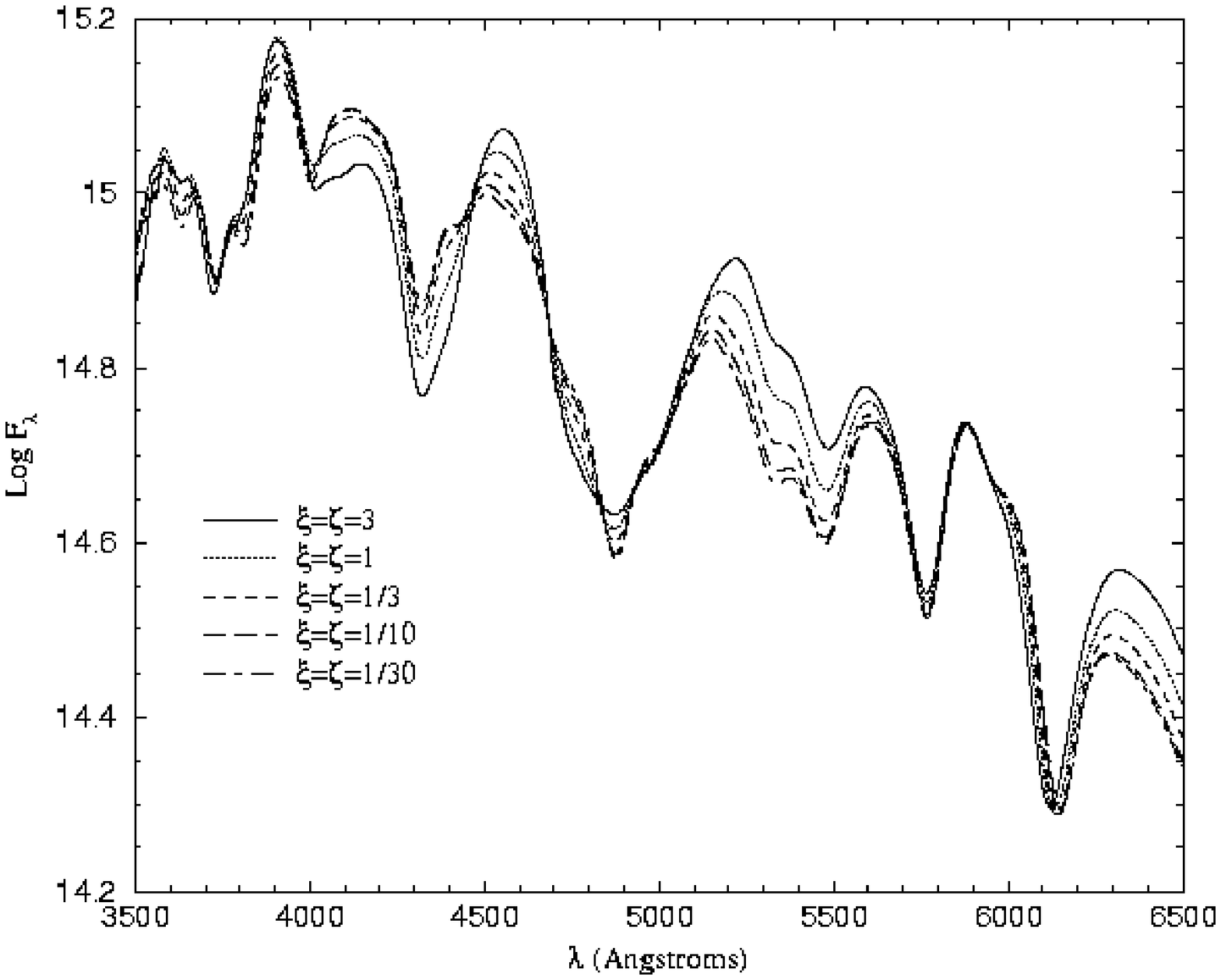,width=14cm,angle=270}
\caption{\label{fig:20fezopt} Expansion of optical region of
Figure~\ref{fig:20fezall}}
\end{center}
\end{figure}

Our grid of synthetic spectra of W7 on day 20 with incomplete burning
zone \feff\ abundance variations and C+O layer metallicity are shown in
Figure~\ref{fig:20fezall}. The effects of the combined 
modifications mostly separate into the effects seen in
Figures~\ref{fig:20zall} \&~\ref{fig:20feall}.  The UV displacements in flux
and the effects on the Si~II 6150~\AA\ feature
in the combined models reflect the
effects of metallicity variation in the C+O layer alone. The
S~II~``W'' at 5300~\AA\ 
and the ``peak erosion'' line at 4100~\AA\ show the primary effect of
\feff\ abundance on the  optical spectra
(Figure~\ref{fig:20fezopt}) when both the \feff\ abundance and the C+O layer metallicity 
 are varied simultaneously.  The effect of the combined
modifications 
on the Ca~II H+K feature are small.  The 4350~\AA\ 
feature in the combined modification has the combination effects of
deeper red wing strength with increasing \feff\ abundance and a deeper
blue wing (Mg~II) with increasing C+O layer metallicity.   The
changes in the C+O layer metallicity and in the \feff\
abundance of the incomplete burning zone each have effects which are
separate from 
one another and combined effect is essentially the sum of the two 
effects.

\subsection{Day 35\label{fe35}}

\begin{figure}
\begin{center}
\leavevmode
%\epsscale{0.8}
%\plotone{d35feall.eps}
\psfig{file=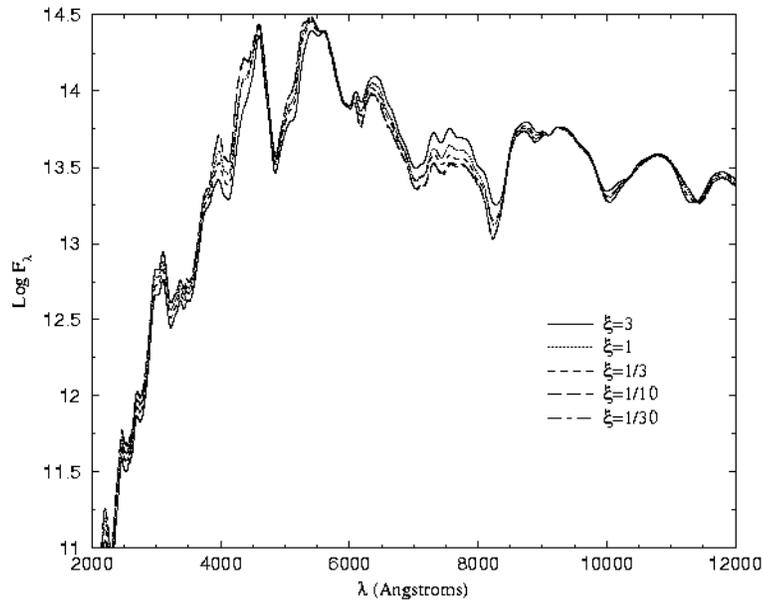,width=14cm,angle=270}
\caption{\label{fig:35feall} Models with various \feff\ abundances in the incomplete
burning layer at 35 days after explosion.
%Solid denotes 3 times the normal \feff\ abundance in the incomplete
%burning zone, thick dotted---normal, short dashed------1/3, long
%dashed---1/10, and dot dashed---1/30.} 
Labels are the same as in Figure~\ref{fig:20feall}.}
\end{center}
\end{figure}

Our grid of synthetic spectra of W7 on day 35 with incomplete 
burning zone \feff\ abundance variations are shown in Figure~\ref{fig:35feall}.
The UV pseudo-continuum varies due to the `surface' cooling and
additional line blocking in the 
incomplete burning zone.   The O~I feature at 8000~\AA\ and
the Si~II feature at 6150~\AA\  become weaker as oxygen 
and silicon are displaced by \feff.  There are several significant effects
in the infrared, however, observations which extend far enough into
the IR to compare with real \snia\ are not available.

\begin{figure}
\begin{center}
\leavevmode
%\epsscale{0.8}
%\plotone{d35fezall.eps}
\psfig{file=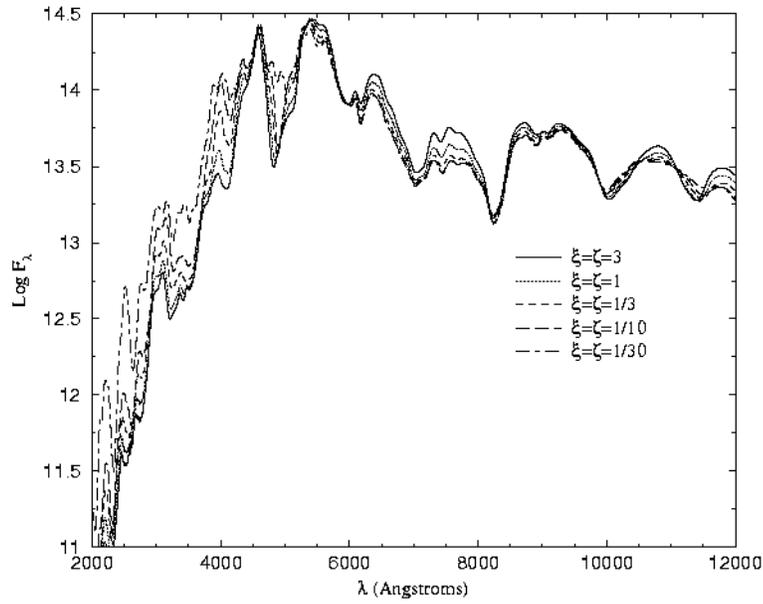,width=14cm,angle=270}
\caption{\label{fig:35fezall} Models with various \feff\ abundances in the incomplete
burning layer and metallicities in the C+O layer at 35 days after explosion.
These models combine the effects in the models in
Figures~\ref{fig:35zall} \&~\ref{fig:35feall}.
%Solid denotes 3 times the normal \feff\ abundance in the incomplete
%burning zone and C+O layer metallicity, thick dotted---normal, short dashed------1/3,
%long dashed---1/10, and dot-dashed---1/30.}
Labels are the same as in Figure~\ref{fig:20fezall}.}
\end{center}
\end{figure}

Our grid of synthetic spectra of W7 on day 35 with combined incomplete
burning zone \feff\ abundance and C+O layer metallicity  variations are shown in
Figure~\ref{fig:35fezall}.  The effects of the combined modifications
mostly separate in the the effects shown in Figures~\ref{fig:35zall}
\&~\ref{fig:35feall}.  The UV pseudo-continuum variation, changes near the
5000~\AA\ Fe~II feature, and the changes to the 11000\AA\ feature
reflect the C+O metallicity modification alone.  The small shift in the
Si~II feature at 6150~\AA\ and the surrounding pseudo-continuum show the
contributions of \feff\ abundance modification.

\section{Discussion}

\subsection{Evolution of S\lowercase{i}~II\label{siii}}

\begin{figure}
\begin{center}
\leavevmode
%\epsscale{0.8}
%\plotone{sivel.eps}
\psfig{file=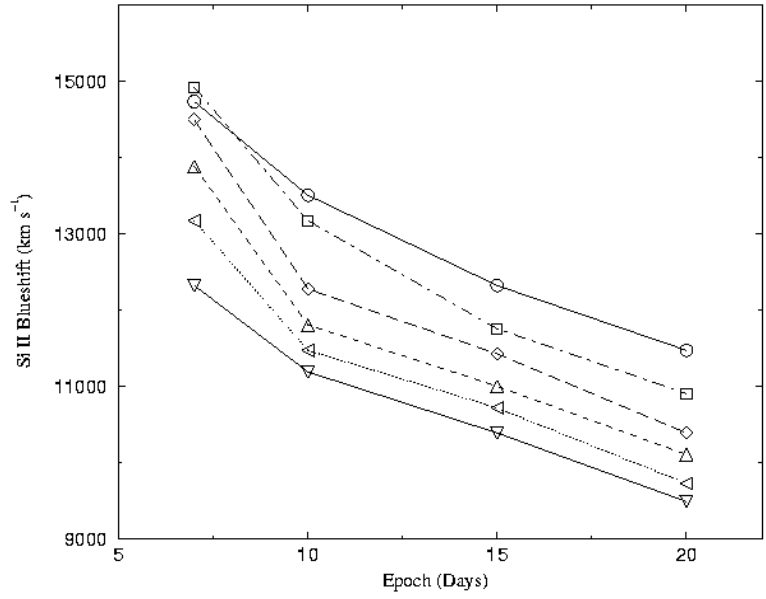,width=14cm,angle=270}
\caption{\label{fig:sivel} Blueshift velocities of Si~II 6150~\AA\ feature
as a function of epoch.  Down triangle symbols represent 1/30  normal
C+O layer metallicity, left triangles 1/10 normal, up triangles
1/3 normal,
diamonds normal, squares 3 times normal, and circles 10 times normal
metallicity. 
Data are from lines shown in Figure~\ref{fig:sigrid}.}
\end{center}
\end{figure}

Figure~\ref{fig:sigrid} shows the evolution of the Si~II 6150~\AA\
feature to maximum light.  The feature grows stronger and steeper as 
line formation of  the Si~II $\lambda6355$ feature  moves into
the silicon rich layers of W7.  As line formation moves into the
silicon rich zones, the effects 
of C+O layer metallicity on line formation, are reduced, but not
eliminated.  The blue-shift velocities 
of the deepest points of the Si~II feature are plotted for these models in
Figure~\ref{fig:sivel}.  Except for the extreme case, $\zeta=10$, the
blue-shift velocities increase monotonically 
with C+O layer metallicity through day
20.  The increasing opacity from the C+O layer
moves the feature blueward.   The velocity shifts
due to
metallicity are degenerate with changes that could be expected from
silicon rich material extending to higher velocities.  However, the effects
of primary line formation in the C+O versus silicon rich layers can be
distinguished by line shapes. These general trends and spreads in
blueshift velocities are similar to those seen in the  data by
\citet{bvdb93}.  They found 
that the slower blueshift velocities tended to be found in earlier
galaxy types. A similar study correlating blueshift velocities, peak
magnitude (or a suitable proxy), and a more quantitative estimate of the
pre-supernova environment metallicity will be the subject of future work.

\subsection{Other Features and Effects}

%There are several effects that evolve with time.  
The UV pseudo-continuum shows the effects of metallicity through the
surface cooling of the C+O layer, additional line blocking and the
shifting of lines which form at 
faster moving layers with higher metallicity.  The UV pseudo-continuum
displacement  is
relatively constant and still present at 35 days after explosion.  The
UV displacement over the entire  range of models is typically
$\sim 0.5$ dex.
 For the more likely range of metallicities, 1/3 to 3 times
solar, the change is up to $\sim 0.2$ dex or 0.5 magnitude.
Unfortunately, we cannot use near-UV flux as a metallicity indicator.  
The UV flux is diagnostic of the temperature, density, and radius
(velocity) of the C+O layer, but is not uniquely determined for any
one quantity.  The 
related backwarming causes pseudo-continuum shifts in the optical and near
infra-red for the early epoch spectra.  This backwarming shift in
continua fades in strength as pseudo-continuum formation moves to deeper
layers. A smaller surface cooling/backwarming effect exists in the
partially burned layers from changes in \feff\ abundance, beginning at
day 20.

\citet{hwt98} also report a change in UV pseudo-continuum for models with
different metallicity by noting decreases in the U band magnitude with
increasing metallicity.  They show a change in flux in the UV spectra
presented,  
but the change occurs in the opposite direction with
metalicity to that which we find here. This is due to the
difference in the density structures between delayed
detonation models (DDs) they employ and that of the
parameterized deflagration model W7 that we use here.
In the DD model the lower metallicity model forms the UV
pseudo-continuum at a smaller radius and so the flux
is lower for smaller metallicities, since the radial
effect wins out over the opacity. 

Some features which had significant impact from C+O
metallicity at early epochs are nearly unaffected at later epochs.  At
day 7 we find that most of the optical features nearly disappear when
the metal content is dropped to 1/30 normal.  This indicates that much
of line formation for these features takes place in the C+O layer.  In
later epochs the impact of the C+O layer metallicity on most optical features
becomes quite small, since line formation occurs mostly in the
deeper layers.

The influence of \feff\ abundance in the incomplete burning zone is very
small at early epochs, since significant
spectrum formation occurs in the C+O layer.    At maximum light and later,
\feff\ abundance variations 
change Fe feature strengths, have some small temperature based effects
on the pseudo-continuum, and change the strength of certain features from
species, e.g. sulfur, which are displaced by \feff\ abundance changes.

%The final feature we discuss is the Ca~II `split'.  
We have confirmed
the \citet{nughydro97} identification of Si~II $\lambda 3858$ as a
component of the Ca~II `split'.  To the blue of the Ca~II H+K feature
is the emission peak from another Ca~II line.  When the Si~II line
forms, the Si~II absorption falls on that peak and the emission from the
Si~II line forms a peak in the center of the `split'.  For models with a
strong Si~II feature, the absorption minimum of the red wing of the Ca~II H+K 
will seem to disappear. 
For W7,
changes in metallicity alter the strength of the Si~II line and thus, the
shape of the Ca `split' feature. In real \snia, other factors may
prevent the formation of a `split' such as stronger Ca~II H+K, weaker
Si~II, or temperature effects.  \citet{jhaetal98bu99} show spectra for
several normal \snia. The Ca~II H+K feature in the observations
display a range of morphologies similar to those seen in our synthetic
spectra. We do not require any abundance or
ionization effect to produce this feature.
While this 
interpretation is strictly correct in the model W7, we suspect it
also produces the observed feature in SNe Ia, however, detailed
comparisons of the calculated velocities to those of the observed
features are required.

\section{Conclusions}

By calculating a series of model atmospheres with abundance variations
around the base W7 model for \snia, we have demonstrated unexpected and
complex effects on the output spectra.  The UV spectra show lower flux
and blueshifting lines as surface cooling, additional line blocking,
and outward movement of the 
line-forming region occur with higher metallicity in the unburned C+O layer. 
We have demonstrated, at epochs well before maximum light that line
formation occurs largely in the C+O layer for species that will later form in
newly synthesized material.  The `splitting' of the Ca~II H+K feature we
can now better understand as a blend of a  Si~II feature with the stronger
Ca~II lines, without abundance or ionization effects in the W7
model. We have shown that the strength, profile, and 
velocity of the \snia\ characteristic Si~II 6150~\AA\ feature are
affected by C+O layer metallicity.  This provides a mechanism for the
variation in the 
blueshift of the Si~II feature without variations in the explosion energy.
For exploding white 
dwarfs of different metallicities \citet{hwt98} and \citet{iwamoto99}
find changes in 
nucleosynthesis.  We tested these effects on spectra by varying
the \feff\ abundance in the 
incomplete burning zone.  We have found that the \feff\ abundance has
negligible effect on pre-maximum spectra and relatively little effect
afterwards.  The  effects of progenitor metallicity variations can mostly be
separated into effects due to \feff, and those due to C+O
metallicity.

Previous spectral studies
\citep{branch81b85,harkness91a,harkness91b,nug1a95,nughydro97} have
shown that W7 has the appropriate 
composition structure to reproduce photospheric era spectra. In
preparing the base fits for this paper we have found that the quality
of the fits decreases away from maximum light. The temperature
structure inversion in the lowest metallicity model at day 35
demonstrates the importance of changes ($\approx 20$\%) in the
temperature structure. We see that density/temperature structures are
important in fitting spectral features. In forthcoming work we will
present the results of numerical experiments to alter the density
structure of W7 to make the model better correspond to the
observations. 

We have shown through parameterized abundance modifications of the
\snia\ model W7 that pre-explosion metallicity can have detectable
effects on the output spectra at every epoch. Due to the uncertainty in
hydrodynamic models and severe blending of lines in \snia\ spectra, we
cannot give a prescriptive analysis tool for measuring the pre-explosion
metallicity of \snia.  However, hopefully this can contribute to the
understanding of the diversity of \snia, and the ways that various
progenitor metallicity effects can affect \snia.  Studies computing 
detailed spectra of hydrodynamic models that include progenitor
metallicity in the evolution of the star and the hydrodynamics and
nucleosynthesis of the supernova should show what effects overall
progenitor metallicity have in creating \snia\ diversity.

\acknowledgments  This work was supported
in part by NSF grants  AST-9731450 and AST-9417102; NASA
grant NAG5-3505, an IBM SUR grant to the University of Oklahoma; and
by NSF grant AST-9720704, NASA ATP grant
NAG 5-3018 and LTSA grant NAG 5-3619 to the University of Georgia.
Some of the calculations presented in this paper were
performed at the the San Diego
Supercomputer Center (SDSC), supported by the NSF, and at the National
Energy Research Supercomputer Center (NERSC), supported by the
U.S. DOE. We thank both these institutions for a generous allocation
of computer time.

\bibliography{refs,sn1bc,sn1a,snii,nucleosynthesis,gals,crossrefs}

\clearpage

\clearpage

\clearpage

\clearpage

\clearpage

\clearpage

\clearpage

\clearpage

\end{document}